\documentclass[twocolumn]{revtex4}
\usepackage{amsmath,amssymb,mathrsfs,url,graphicx}

\newcommand{\be}{\begin{equation}}
\newcommand{\ee}{\end{equation}}
\newcommand{\Hilbert}{\mathscr{H}}

\newcommand{\RRR}{\mathbb{R}}
\newcommand{\CCC}{\mathbb{C}}
\newcommand{\scp}[2]{\langle #1|#2 \rangle}
\newcommand{\Bscp}[2]{\Bigl\langle #1\Big|#2 \Bigr\rangle}

\newcommand{\vq}{\boldsymbol{q}}
\newcommand{\vx}{\boldsymbol{x}}
\newcommand{\vy}{\boldsymbol{y}}
\newcommand{\vz}{\boldsymbol{z}}
\newcommand{\M}{\mathcal{M}}

\begin{document}

\title{Matter Density and Relativistic Models of Wave Function Collapse}
   
\author{Daniel Bedingham}
\email{d.bedingham@imperial.ac.uk}
\affiliation{Blackett Laboratory, Imperial College, London~SW7~2BZ, 
	United Kingdom}
	 
\author{Detlef D\"urr}
\email{duerr@mathematik.uni-muenchen.de}
\affiliation{Mathematisches Institut, Ludwig-Maximilians-Universit\"at, 
	Theresienstr.~39, 80333~M\"unchen, Germany}

\author{GianCarlo Ghirardi}
\email{ghirardi@ts.infn.it}
\affiliation{Department of Theoretical Physics of the University of Trieste, 
	and the Abdus Salam International Centre for Theoretical Physics, Trieste, 
	and Istituto Nazionale di Fisica Nucleare, Sezione di Trieste, Italy}

\author{Sheldon Goldstein}
\email{oldstein@math.rutgers.edu}
\affiliation{Departments of Mathematics, Physics and
     Philosophy, Rutgers University, Hill Center,  
     110 Frelinghuysen Road, Piscataway, NJ 08854-8019, USA}

\author{Roderich Tumulka}
\email{tumulka@math.rutgers.edu}
\affiliation{Department of Mathematics,
     Rutgers University, Hill Center, 
     110 Frelinghuysen Road, Piscataway, NJ 08854-8019, USA}

\author{Nino Zangh\`\i}
\email{zanghi@ge.infn.it}
\affiliation{Dipartimento di Fisica dell'Universit\`a
     di Genova and INFN sezione di Genova, Via Dodecaneso 33, 16146
     Genova, Italy}
     
\date{June 7, 2013}

\begin{abstract}
Mathematical models for the stochastic evolution of wave functions that combine the unitary evolution according to the Schr\"odinger equation and the collapse postulate of quantum theory are well understood for non-relativistic quantum mechanics. Recently, there has been progress in making these models relativistic. But even with a fully relativistic law for the wave function evolution, a problem with relativity remains: Different Lorentz frames may yield conflicting values for the matter density at a space-time point. We propose here a relativistic law for the matter density function. According to our proposal, the matter density function at a space-time point $x$ is obtained from the wave function $\psi$ on the past light cone of $x$ by setting the $i$-th particle position in $|\psi|^2$ equal to $x$, integrating over the other particle positions, and averaging over $i$. We show that the predictions that follow from this proposal agree with all known experimental facts.
\end{abstract}



\maketitle

\begin{center}
{\it Dedicated to Herbert Spohn\\ on the occasion of his 65th birthday}
\end{center}

\section{Background}

It is widely believed that in our quantum world physical facts about
occurrences in space-time must be grounded in the wave function. There
are several long recognized and much discussed difficulties with this
view:
\begin{enumerate}
\item To the extent that one can discern these facts in the wave function,
they seem to originate primarily (if not exclusively) in wave function
collapse upon measurement, and not in the fundamental wave function
dynamics itself, given by Schr\"odinger's equation, which appears to be
incompatible with collapse.

\item The nature of these facts is rather nebulous, in part because the
notion of measurement is itself rather vague, but also because no precise
specification is provided for the connection between the wave function
and clear structures in space-time.

\item The facts that originate via collapse upon measurement seem very much
to depend upon the choice of Lorentz frame \cite{Blo67}, and thus
seem to conflict with special relativity.
\end{enumerate}

One way to deal with the first problem involves replacing Schr\"odinger's
equation with a suitable stochastic evolution law for the wave function (a ``law for $\psi$'') that remains close to the unitary Schr\"odinger evolution for systems consisting of few particles but avoids superpositions of macroscopically different states (such as Schr\"odinger's cat) for macroscopic systems. This is the approach of \emph{spontaneous-collapse theories} \cite{Ghi07} (such as the GRW theory \cite{GRW86,Bell87} and the CSL theory \cite{Pe89}, a.k.a.\ dynamical state reduction theories; hereafter, ``collapse theories''), also considered by \cite{Dio89,GP93,Leg02,Pen00,Adl07,Wein11}, among others.

As for the second problem, Ghirardi et al.\ \cite{BGG95} have proposed
that we define a ``matter density'' in terms of the wave function;Ê a
simple example of such a specification will be given shortly.

To deal with the third problem, Hellwig and Kraus (HK) \cite{HK70}, following Bloch \cite{Blo67}, proposedÊ
that a measurement at space-time point $x$ lead to a collapse along the
past light cone of $x$. For example, for a single-particle system a
detection at $x$ should imply the vanishing of its wave function everywhere
in space-time except within the light cone (past and future) of $x$.
However, Aharonov and Albert (AA) \cite{AA80,AA81}, following Landau and Peierls \cite{LP31}, have effectively argued that the proposal of HK is very problematical. Following Bloch,
they insist that upon measurement the effect of the associated collapse
must be allowed to depend upon the choice of Lorentz frame. More
generally, the wave function must be regarded, according to AA, 
as depending upon a choice of spacelike hypersurface,Ê with
differentÊ surfacesÊ that locally agree sometimes supportingÊ wave
functions---and hence facts---that are respectively entirely different,
even locally. For example, consider an EPR experiment \cite{Bell87b}, in which two particles in the singlet spin state are widely separated in space, and a Stern--Gerlach experiment is carried out on each particle. The reduced spin state $\rho$ of particle 1 (obtained by tracing out the spin of particle 2) will depend on the choice of hypersurface $\Sigma$: If $\Sigma$ lies after the experiment on particle 2 but before that on particle 1, then $\rho$ will be a pure state. If $\Sigma$ lies before both experiments, $\rho$ will be mixed. 
The problem of finding a consistent relativistic
specification of facts thus remains.

\section{Scope}

In this paper we describe a theory that resolves all the puzzles
revolving around relativistic wave function collapse. It does this by
combining aspects of the views of HK and AA to provideÊ a fully
relativistic version of spontaneous localization involving as a basic
variable a covariant specification of matter density on the microscopic
level. The theory follows AA in that it involves a wave function
evolution (with spontaneous collapses) that associates a wave function
with every spacelike hypersurface (as well as with past light cones).
Following HK, past light cones nonetheless play a distinguished role:
Not, as with HK, to define the collapse of the wave function, but rather
to define the matter density. In our theory, the matter density at a
space-time point $x$ is defined in terms of the wave function associated
with the past light cone of $x$. In this way we obtain a fully relativistic
collapse theory that, like the GRW theory,
reproduces the quantum predictionsÊ for all experiments performed so far.
We do not aim at mathematical rigor and leave aside various technical subtleties such as regularization procedures.

\section{Matter density}

For an interpretation of quantum theory to be satisfactory, we should demand that certain local facts, such as whether a cat is dead or alive, do not depend on the choice of  hypersurface. Fortunately, the \emph{macroscopic} local situation is practically unambiguous in relevant collapse theories. But since the notion of ``macroscopic'' is inevitably imprecise, a satisfactory version of a collapse theory needs to introduce some variables defining local facts also on the \emph{microscopic} scale. The variable that defines local facts need not define a spin state for every particle. But it should define the distribution of matter in space-time and ensure that macroscopic configurations (e.g., positions of pointers of measurement instruments) are unambiguous; the technical name for this variable is ``the primitive ontology'' (PO). At least two different choices of PO have been suggested for collapse theories: ``flashes'' (e.g., \cite{Bell87,Tum06,AGTZ06}) and the ``matter density function'' (e.g., \cite{Dio89,BGG95,Gol98,AGTZ06}). According to the latter choice, matter is continuously distributed in space with matter density function
\be\label{m}
m(\vx,t) = \sum_{i=1}^N m_i \int_{\RRR^{3N}} d^{3N}\!q\; \delta^3(\vx-\vq_i) \, |\psi_t(q)|^2\,,
\ee
where $q=(\vq_1,\ldots,\vq_N)$ is the configuration variable of a non-relativistic $N$-particle universe, $\psi_t$ is the wave function of this universe at time $t$ (as obtained from the stochastic law for $\psi$), and $m_i$ is the constant usually called the mass of particle number $i$. We call Eq.~\eqref{m} a ``law for $m$.'' Eq.~\eqref{m} can be reformulated as $m(\vx,t) = \scp{\psi_t}{\M(\vx)|\psi_t}$, where $\M(\vx)$ is the mass density operator at $\vx\in\RRR^3$, defined on $L^2(\RRR^{3N})$ to be the multiplication operator $\M(\vx) = \sum_{i=1}^N m_i \, \delta^3(\vx-\vq_i)$.

Since matter is distributed continuously, there are strictly speaking no particles according to this theory. Nevertheless, we will often adopt conventional language and speak of particles, thereby meaning the variables $\vq_i$ in the wave function. It is known that with the PO given by (either flashes or) a matter density function, the empirical predictions of the GRW and CSL theories deviate from those of standard quantum theory only so slightly that experimental tests distinguishing them from standard quantum theory are not possible to date \cite{Leg02,JPR04,Ghi07,Adl07,FT12}, though possible in principle.
 
Eq.~\eqref{m} is not Lorentz invariant because it involves integrating over all positions $\vq_i$ \emph{at the same time} $t$, so that for any space-time point $x$ the value of $m(x)$ depends on the Lorentz frame chosen for evaluating Eq.~\eqref{m}.

\section{Relativistic law for $\psi$}

Progress to date towards a relativistic law for $\psi$ involving spontaneous collapse includes: a toy model providing a Lorentz-invariant process for $\psi$ in terms of given ``measurement'' events \cite[Sec.~7]{Ghi}; relativistic processes for the state vector of a quantum field theory \cite{Dio90,Pea90,GGP90} which, however, suffer from divergences; a recent modification of one of these processes avoiding the divergences \cite{Bed10}; and a relativistic GRW process for the state vector of $N$ non-interacting spin-$\frac12$ particles in an external field \cite{Tum06}. 

Our scheme works with any of these laws for $\psi$; we only use that the law for $\psi$ has the properties (P1)--(P4) below. We use the Schr\"odinger picture in the Tomonaga--Schwinger variant, in which with every spacelike hypersurface $\Sigma$ there is associated a vector $\psi_\Sigma$ in a Hilbert space $\Hilbert$ with $\|\psi_\Sigma\|=1$. It will sometimes be convenient to work in a representation using a separate Hilbert space $\Hilbert_\Sigma$ for every $\Sigma$, with $\psi_\Sigma\in\Hilbert_\Sigma$. For simplicity, we assume that the world history begins on a spacelike hypersurface $\Sigma_0$, say at a big bang singularity; however, this assumption is not indispensable. For any space-time point $x=x^\mu=(\vx,t)$, let $PLC(x)$ denote the past light cone of $x$, more precisely the hypersurface formed by the past light cone of $x$ (down to its intersection with  $\Sigma_0$) together with the part of $\Sigma_0$ outside the past of $x$. We use that:
\begin{itemize}
\item[(P1)] Given an initial wave function $\psi_0$ on $\Sigma_0$ (and possibly further data), the law for $\psi$ specifies the joint distribution of all $\psi_\Sigma$ with $\Sigma$ in the future of $\Sigma_0$.
\item[(P2)] Apart from spacelike hypersurfaces, the $\Sigma$ in (P1) can also be $PLC(x)$ for any space-time point $x$.
\item[(P3)] In situations in which the unitary Schr\"odinger evolution would lead to a superposition $\psi_\Sigma=\sum_\alpha c_\alpha \psi^{(\alpha)}$ of macroscopically different contributions $\psi^{(\alpha)}$ (with $\|\psi^{(\alpha)}\|=1$), the law for $\psi$ yields $\psi_\Sigma\approx \psi^{(\alpha)}$ with probability close to $|c_\alpha|^2$.
\item[(P4)] For any two hypersurfaces $\Sigma,\Sigma'$ after a local measurement at a space-time point $y$, $\psi_\Sigma$ and $\psi_{\Sigma'}$ select the same $\alpha$ of that measurement (using the notation of (P3)). 
\end{itemize}
Property (P3) is a basic feature of collapse theories \cite{Ghi07}. For the models of \cite{Pea90,GGP90,Ghi,Tum06,Bed10}, (P1), (P3), and (P4) are true. Property (P2) also holds whenever the unitary evolution (without collapse) extends to lightlike hypersurfaces; although lightlike hypersurfaces are often not Cauchy hypersurfaces, this is known to be true \cite{Tum09} at least for systems (with suitable potentials) containing no massless particles.

\section{Relativistic law for $m$}

We propose the following relativistic law for $m$, which was outlined in \cite{Tum07,Bed10} using ideas from \cite{Ghi99}:
\be\label{rmM}
m(x) = \Bscp{\psi_{PLC(x)}}{\M_{PLC(x)}(x)\Big|\psi_{PLC(x)}}\,,
\ee
where $\scp{\cdot}{\cdot}$ is the inner product in $\Hilbert_{PLC(x)}$, and $\M_{PLC(x)}(x)$ is the mass density operator at $x$ in $\Hilbert_{PLC(x)}$. If $\M_{PLC(x)}(x)$ is a 4-vector or tensor, then so is $m(x)$. We shall call a theory such as described above, based on \eqref{rmM} and (P1)--P(4), a
{\em relativistic matter density based collapse theory}.

As a concrete example, if the law for $\psi$ is a collapse version of a quantum field theory then $\M(x)$ should be the stress-energy-momentum-tensor operator $\mathcal{T}_{\mu\nu}(x)$. Note that while for a conventional quantum field theory, i.e., one without spontaneous collapses and with a unitary time evolution, the expression $\scp{\psi_\Sigma}{\mathcal{T}_{\mu\nu}(x)|\psi_\Sigma}$ is independent of the choice of $\Sigma$, the expression does depend on $\Sigma$ in collapse theories (because when changing $\Sigma$ to $\Sigma'$, the collapses during the evolution from $\Sigma$ to $\Sigma'$ may affect its value)---so it is relevant to specify that $\Sigma=PLC(x)$.

As another concrete example, consider a universe consisting of $N$ Dirac particles. Then $\Hilbert_\Sigma$ consists of those functions $\psi:\Sigma^N\to (\CCC^4)^{\otimes N}$ with
\be
\int_{\Sigma^N} d\sigma^{\mu_1}(y_1) \cdots d\sigma^{\mu_N}(y_N) \: \overline{\psi} [\gamma_{\mu_1}\otimes \cdots \otimes \gamma_{\mu_N}] \psi<\infty\,,
\ee
where $\psi$ is evaluated at $(y_1,\ldots,y_N)\in\Sigma^N$ and the vector-valued measure $d\sigma^\mu(y)$ on $\Sigma$ is defined by the vector-valued differential 3-form obtained from the invariant 4-form $\varepsilon_{\kappa\lambda\mu\nu}$ by raising the first index. (For spacelike $\Sigma$, $d\sigma^\mu(y) = n^\mu(y) \, d^3y$ with $n^\mu(y)$ the future-pointing unit normal vector on $\Sigma$ at $y$ and $d^3y$ the volume in the sense of the 3-metric on $\Sigma$. For $\Sigma=PLC(x)$, $d\sigma^\mu(y)$ is the Lorentz-invariant vector-valued measure on $PLC(x)$ obtained as the product of the invariant vector field $x^\mu-y^\mu$ on $PLC(x)$ and the invariant scalar measure with coordinate expression $dy^1 \, dy^2\, dy^3/(x^0-y^0)$.)

In this case, $m(x)=m_\mu(x)$ is the 4-vector field
\begin{multline}
m_\mu(x) =\sum_{i=1}^N m_i \delta_{\mu}^{\mu_i} \int_{PLC(x)^{N-1}} \Bigl( \prod_{j\neq i} d\sigma_j^{\mu_j}(y_j) \Bigr)\:\times \\
\times \: \overline{\psi}_{PLC(x)} [\gamma_{\mu_1}\otimes \cdots \otimes \gamma_{\mu_N}] \psi_{PLC(x)}\,,
\end{multline}
with $\psi$ evaluated at $(y_1,\ldots,y_{i-1},x,y_{i+1},\ldots,y_N)$.

The law for $m$, Eq.~\eqref{rmM}, is relativistic and does not invoke any notion of simultaneity of spacelike separated points. In the \emph{non-relativistic limit}, formally $c\to\infty$, the law approaches \eqref{m}, as $PLC(x)$ approaches a hyperplane of constant time (viz., $t=x^0$) in that limit.

\section{Quantum measurements}

Assertions  \eqref{claim5} and \eqref{claim6} below are the crucial observations concerning the connection between the predictions of a relativistic matter density based  collapse theory and those of standard quantum theory.
\be\label{claim5}
\begin{minipage}{70mm}
Suppose that a local measurement is made at a space-time point $y$ as described in (P4).  Then in the future light cone of $y$, the $m$ function is one in which the apparatus pointer points to $\alpha$. That is, the $m$ function and the $\psi$ function (on any $\Sigma$ after $y$) agree about the outcome.
\end{minipage}
\ee
Proof: According to the law for $m$, Eq.~\eqref{rmM}, the values $m(x)$ at points $x$ in the future light cone of $y$ representing the pointer are determined by $\psi_{PLC(x)}$. By (P3) and (P4), $\psi_{PLC(x)}$ is (approximately) collapsed to one outcome $\alpha$. By (P4), that $\alpha$ is independent of $x$. By \eqref{rmM}, the $m$ function is near zero if $x$ is a location at which the pointer would indicate an outcome $\neq\alpha$ but substantial if $x$ is a location indicating $\alpha$. Thus, the pointer (made out of the $m$ function) points to $\alpha$. By (P4), $\psi_\Sigma$ on any $\Sigma$ after $y$ selects the same outcome $\alpha$.

As a consequence of \eqref{claim5}, the outcome (defined by the $m$ function!)~can be read off from $\psi_\Sigma$ on a hypersurface $\Sigma$ after $y$, notwithstanding our insistence that the pointer consists of $m$, not of $\psi$. As a further consequence of \eqref{claim5} and (P3), the probabilities of different pointer positions agree (approximately) with the probabilities assigned to the different outcomes by the quantum formalism. The same is true of the joint probability distribution of several pointers reporting the outcomes of several local measurements at $y_1,\ldots,y_n$, as follows by applying \eqref{claim5} to each of the $y_i$ and to a hypersurface $\Sigma$ after all of the $y_i$. Since any experiment ultimately consists of local measurements, we obtain that 
\be\label{claim6}
\begin{minipage}{70mm}
The empirical predictions of a relativistic matter density based collapse theory agree (approximately) with those of the quantum formalism. 
\end{minipage}
\ee

It follows from \eqref{claim6} by virtue of Bell's theorem \cite{Bell87b} that collapse theories with our law for $m$, Eq.~\eqref{rmM}, although relativistic, are \emph{non-local}, i.e., there is faster-than-light action-at-a-distance.

\section{Example}

\begin{figure}[ht]
\begin{center}
\includegraphics[width=70mm]{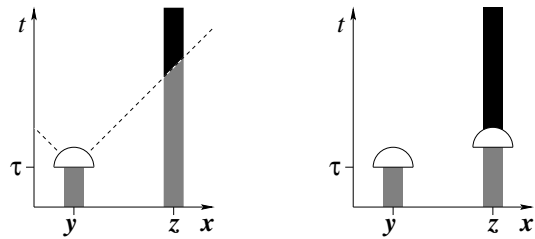}
\caption{\label{fig:ex}Space-time diagram of $m(x)$ in the example. Semi-circles represent detectors, dashed lines a light cone, thick lines the region where $m_1(x)=m_0$ (black) or $m_1(x)=m_0/2$ (grey). LEFT: with one detector, RIGHT: with two detectors.}
\end{center}
\end{figure}

Consider a single particle and a detector. At time $t=0$ in some Lorentz frame, let the wave function be
\be\label{expsi}
\psi_0 = \frac{1}{\sqrt{2}} \Bigl( |\vy\rangle + |\vz\rangle \Bigr) |\text{ready}\rangle\,,
\ee
where $|\vy\rangle$ and $|\vz\rangle$ are well-localized wave packets centered at (distant) space points $\vy$ and $\vz$, respectively, and $|\text{ready}\rangle$ is a state of the environment in which the detector is ready. Suppose that, in that Lorentz frame, the detector is at rest at $\vy$, and the interaction between the particle and the detector is turned on at time $\tau>0$. Then $\psi_\Sigma$ has collapsed to either $\approx|\vy\rangle|\text{fired}\rangle$ or $\approx|\vz\rangle|\text{not fired}\rangle$ for every hypersurface $\Sigma$ after the space-time point $(\vy,\tau)$. Consider the latter possibility. Then the contribution $m_1(x)$ to the $m(x)$ function from the single particle according to Eq.~\eqref{rmM} is as depicted in Fig.~\ref{fig:ex} left, with suitable constant $m_0$.

That is, the change in $m$ at $\vz$ is delayed at the speed of light. As a consequence, there is no conservation law for $\int_\Sigma m(x) \, d^3x$---this quantity depends on $\Sigma$, and that is not a problem. Readers might worry, however, that because of the delay another detector at $(\vz,\tau+\varepsilon)$ might have a probability of only $1/2$ (instead of 1) to be triggered. This is not so, as is obvious from \eqref{claim6}. More directly from (P4), for any hypersurface $\Sigma$ after both $(\vy,\tau)$ and $(\vz,\tau+\varepsilon)$, $\psi_\Sigma$ has collapsed corresponding to exactly one detector having fired. The function $m_1(x)$ changes accordingly at each detection (Fig.~\ref{fig:ex} right).

\section{EPR-type experiments}

\begin{figure}[ht]
\begin{center}
\includegraphics[width=85mm]{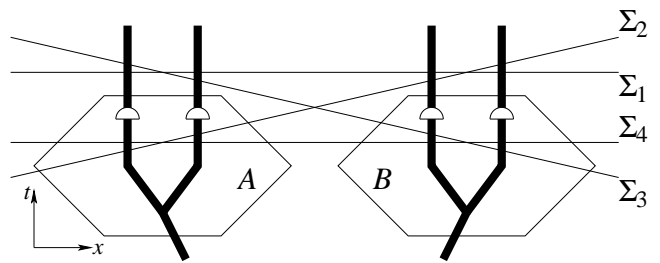}
\caption{\label{fig:EPR1}Space-time diagram of EPR experiment together with some hypersurfaces; semicircles represent detectors; thick lines indicate where the two EPR particles can move; they split when passing the Stern--Gerlach magnets.}
\end{center}
\end{figure}

In these experiments, one performs a Stern--Gerlach experiment on each of two particles in the singlet spin state, one in region $A$ and one in $B$, with $A$ and $B$ spacelike separated; see Fig.~\ref{fig:EPR1}. We shall denote by $O_A, O_B\in \{-1, +1\}$ the outcomes of these  experiments. Let $\Sigma_1$ be a spacelike hypersurface after $A$ and $B$, as in Fig.~\ref{fig:EPR1}, and $\psi$ the wave function of the two EPR particles together with all relevant apparatuses.

We focus on the contribution to $m$ from the two EPR particles themselves. To this end, consider a hypersurface $\Sigma_2$ that passes through $A$ before detection, but after the particle has passed the Stern--Gerlach magnet; and likewise with $\Sigma_3$ in $B$; see Fig.~\ref{fig:EPR1}.  
\be\label{claim3}
\begin{minipage}{70mm}
The $m$ function of the EPR pair restricted to $\Sigma_1$ agrees (approximately) with what would have been obtained by a naive application of Eq.~\eqref{m} in a frame in which $\Sigma_1$ is a constant-time hypersurface.
\end{minipage}
\ee
However:
\be\label{claim4}
\begin{minipage}{70mm}
The statements analogous to \eqref{claim3} for $\Sigma_2$ and $\Sigma_3$ are generally not true.
\end{minipage}
\ee
Proof: Claim \eqref{claim3} is shown in much the same way as \eqref{claim5}, as follows. Let $\psi$ be again the wave function of the EPR pair together with all apparatus, and let us suppose $O_A=+1$. Then both $\psi_{\Sigma_1}$ and $\psi_{PLC(x)}$ for $x$ on the $A$ side of $\Sigma_1$ have so collapsed that the EPR particle on the $A$ side has (say) wave function $|\text{right}\rangle$. (The word ``approximately'' in \eqref{claim3} refers to the fact that the collapsed-away contribution is not exactly zero, just tiny.) Thus, $m(x)=0$ in the left channel and $m(x)=m_0$ in the right channel, in agreement with what would have been obtained from Eq.~\eqref{m} on $\Sigma_1$. The same argument applies to $B$, which proves \eqref{claim3}. To prove \eqref{claim4} for (say) $\Sigma_2$, consider a hypersurface $\Sigma_4$ passing through both $A$ and $B$ just before the detection as in Fig.~\ref{fig:EPR1} and a wave function 
\be\label{EPRpsi}
\psi_{\Sigma_4} = \frac{1}{\sqrt{2}} 
\Bigl( |\text{left}\rangle |\text{right}\rangle - |\text{right}\rangle |\text{left}\rangle \Bigr)|\text{ready}\rangle
\ee
(as would arise from the spin singlet state by choosing the same orientation for both magnets). For $x\in A\cap \Sigma_2$, $m(x)$ is determined by the uncollapsed wave function $\psi_{PLC(x)}$ and thus is $m_0/2$ in the left channel and $m_0/2$ in the right. However, since $\psi_{\Sigma_2}$ is collapsed thanks to the detection in $B$, it is (approximately) $|\text{right}\rangle |\text{left}\rangle$, so Eq.~\eqref{m} on $\Sigma_2$ would yield a matter density on the $A$ side that is $0$ in the left channel and $m_0$ in the right, q.e.d.

\begin{acknowledgments} 
G.C.G.\ and N.Z.\ are supported in part by INFN, Sezioni di Trieste e Genova. 
D.D., S.G., G.C.G., and N.Z.\ are supported in part by the COST-Action MP1006.
R.T.\ is supported in part by NSF Grant SES-0957568 and by the Trustees Research Fellowship Program at Rutgers, the State University of New Jersey. 
S.G.\ and R.T.\ are supported in part by grant no.\ 37433 from the John Templeton Foundation.
\end{acknowledgments}


\begin{thebibliography}{20}

\bibitem{Adl07} S. L. Adler:
  Lower and Upper Bounds on CSL Parameters from Latent Image
  Formation and IGM Heating.
  \textit{J. Phys. A: Math. Theor.}
  \textbf{40}: 2935--2957 (2007).
  arXiv:quant-ph/0605072

\bibitem{AA80} Y. Aharonov, D. Z. Albert: 
	States and observables in relativistic quantum field theories.
	\textit{Phys. Rev. D} {\bf 21}: 3316--3324 (1980)

\bibitem{AA81} Y. Aharonov, D. Z. Albert:
	Can we make sense out of the measurement process in relativistic quantum mechanics?
	\textit{Phys. Rev. D} {\bf 24}: 359--370 (1981)

\bibitem{AGTZ06} V. Allori, S. Goldstein, R. Tumulka, N. Zangh\`\i:
  On the Common Structure of Bohmian Mechanics and the
  Ghirardi--Rimini--Weber Theory. 
  \textit{Brit. J. Philos. Sci.} \textbf{59}: 353--389 (2008).
  arXiv:quant-ph/0603027

\bibitem{Bed10} D. Bedingham: Relativistic state reduction dynamics.
	\textit{Found. Phys.} \textbf{41}: 686--704 (2011). arXiv:1003.2774

\bibitem{Bell87} J. S. Bell: Are There Quantum Jumps? Pages 41--52 in C. W. Kilmister (ed.),
  \textit{Schr\"odinger. Centenary Celebration of a Polymath.} Cambridge
  University Press (1987). Reprinted as chapter 22 of \cite{Bell87b}. 

\bibitem{Bell87b} J. S. Bell: \textit{Speakable and Unspeakable in
    Quantum Mechanics}. Cambridge University Press (1987)

\bibitem{BGG95} F. Benatti, G.C. Ghirardi, R. Grassi: 
	Describing the macroscopic world: closing the circle within
	the dynamical reduction program.
	\textit{Found. Phys.} {\bf 25}: 5--38 (1995)

\bibitem{Blo67} I. Bloch: 
	Some relativistic oddities in the quantum theory of observation. 
	\textit{Phys. Rev.} {\bf 156}: 1377--1384 (1967)

\bibitem{Dio89} L. Di\'osi:
   Models for universal reduction of macroscopic quantum
   fluctuations. \textit{Phys. Rev. A} \textbf{40}: 1165--1174 (1989)

\bibitem{Dio90} L. Di\'osi:
   Relativistic theory for continuous measurement of quantum fields.
   \textit{Phys. Rev. A} \textbf{42}: 5086--5092 (1990)

\bibitem{FT12} W. Feldmann, R. Tumulka:
	Parameter Diagrams of the GRW and CSL Theories of Wave Function Collapse.
	\textit{J. Phys. A} {\bf 45}: 065304 (2012).
	arXiv:1109.6579

\bibitem{Ghi99} G.C. Ghirardi: Some Lessons from Relativistic
   Reduction Models. 
   Pages 117--152 in H.-P. Breuer, F. Petruccione (ed.s), \textit{Open Systems and 
   Measurement in Relativistic Quantum Theory}.
   Berlin: Springer-Verlag (1999)

\bibitem{Ghi} G.C. Ghirardi: Local Measurements of Nonlocal Observables and
	the Relativistic Reduction Process. 
	\textit{Found. Phys.} \textbf{30}: 1337--1385 (2000)

\bibitem{Ghi07} G.C. Ghirardi: Collapse Theories. 
	In E. N. Zalta (ed.), \textit{Stanford Encyclopedia of Philosophy}, 
	published online by Stanford University. 
	\url{http://plato.stanford.edu/entries/qm-collapse/}
	(2007)

\bibitem{GGP90} G.C. Ghirardi, R. Grassi, P. Pearle:
	Relativistic dynamical reduction models: General framework and examples.
	\textit{Found. Phys.} \textbf{20}: 1271--1316 (1990)

\bibitem{GRW86} G.C. Ghirardi, A. Rimini, T. Weber: Unified
  Dynamics for Microscopic and Macroscopic Systems. \textit{Phys. Rev.
    D} \textbf{34}: 470--491 (1986)

\bibitem{GP93} N. Gisin, I. C. Percival:
        The quantum state diffusion picture of physical processes.
        \textit{J. Phys. A: Math. Gen.} \textbf{26}: 2245--2260 (1993)

\bibitem{Gol98} S. Goldstein: Quantum Theory Without Observers.
  \textit{Physics Today}, Part One: March 1998, 42--46. 
  Part Two: April 1998, 38--42.

\bibitem{HK70} K.-E. Hellwig, K. Kraus:
	Formal Description of Measurements in Local Quantum Field Theory.
	\textit{Phys. Rev. D} \textbf{1}: 566--571 (1970)

\bibitem{JPR04} G. Jones, P. Pearle, J. Ring:
  Consequence for Wavefunction Collapse Model of the Sudbury
  Neutrino Observatory Experiment.
  \textit{Found. Phys.} \textbf{34}: 1467--1474 (2004).
  arXiv:quant-ph/0411019
  
\bibitem{LP31} L. Landau, R. Peierls: 
	Erweiterung des Unbestimmt\-heits\-prinzips f\"ur die
	relativistische Quantentheorie. 
	\textit{Z. Physik} {\bf 69}: 56--69 (1931)

\bibitem{Leg02} A. J. Leggett: Testing the limits of quantum
   mechanics: motivation, state of play, prospects.
   \textit{J. Phys.: Cond. Matt.} \textbf{14}: R415--R451 (2002)

\bibitem{Pe89} P. Pearle: Combining stochastic dynamical state-vector 
   reduction with spontaneous localization. \textit{Phys. Rev. A} \textbf{39}: 
   2277--2289 (1989)

\bibitem{Pea90} P. Pearle: Toward a relativistic theory of statevector reduction.
	In A. I. Miller (ed.): 
	\textit{Sixty-Two Years of Uncertainty}, 
	volume 226 of \textit{NATO ASI Series B}. New York: Plenum Press (1990)

\bibitem{Pen00} R. Penrose: Wavefunction Collapse As a Real
   Gravitational Effect. Pages 266--282 in A. Fokas, T. W. B. Kibble, A. Grigoriou,
   B. Zegarlinski (ed.s), \textit{Mathematical Physics 2000}.
   London: Imperial College Press (2000)

\bibitem{Tum06} R. Tumulka: A relativistic version of the Ghirardi--Rimini--Weber model.
    \textit{J. Statist. Phys.} \textbf{125}: 821--840 (2006). 
    arXiv:quant-ph/0406094

\bibitem{Tum07} R. Tumulka: The `Unromantic Pictures' of Quantum Theory.
	\textit{J. Phys. A: Math. Theor.} 
	\textbf{40}: 3245--3273 (2007). arXiv:quant-ph/0607124

\bibitem{Tum09} R. Tumulka: The Point Processes of the GRW Theory of Wave Function Collapse.
 	\textit{Rev. Math. Phys.} \textbf{21}: 155--227 (2009). 
	arXiv:0711.0035

\bibitem{Wein11} S. Weinberg: Collapse of the State Vector.
	\textit{Phys. Rev. A} {\bf 85}: 062116 (2012).
	arXiv:1109.6462

\end{thebibliography}
\end{document}